\documentclass[final,5p,times,twocolumn]{elsarticle}

\usepackage{graphics}
 \usepackage{graphicx}
 \usepackage{epstopdf}
 \usepackage{epsfig}
\usepackage{amssymb}
\usepackage{amsmath}
\usepackage{lineno}

\usepackage{tikz}
\usepackage{lipsum}
\usepackage{subfig}
\usepackage{booktabs}

\begin{document}
\begin{frontmatter}
\title{Preliminary Study for an RF photocathode based electron injector for awake project }
\author{\small{O. Mete, G. Xia, The University of Manchester, Manchester, UK and Cockcroft Institute Sci-Tech Daresbury, Warrington, UK,}
\\\small{Graeme Burt, The University of Lancaster, Lancaster, UK and Cockcroft Institute Sci-Tech Daresbury, Warrington UK}
\\ \small{S. Chattopadhyay, The University of Liverpool, UK, The University of Manchester, Manchester, UK,} \\ \small{The University of Lancaster, Lancaster, UK and Cockcroft Institute Sci-Tech Daresbury, Warrington UK}}
\begin{abstract}
AWAKE project, a proton driven plasma wakefield acceleration (PDPWA) experiment is approved by CERN. The PDPWA scheme consists of a seeding laser, a drive beam to establish the accelerating wakefields within the plasma cell; and a witness beam to be accelerated. The drive beam protons will be provided by the CERN's Super Proton Synchrotron (SPS). The plasma ionisation will be performed by a seeding laser and the drive beam protons to produce the accelerating wakefields. After establishing the wakefields, witness beam, namely, electron beam from a dedicated source should be injected into the plasma cell. The primary goal of this experiment is to demonstrate acceleration of a 5-15$\,$MeV single bunch electron beam up to 1$\,$GeV in a 10$\,$m of plasma. This paper explores the possibility of an RF photocathode as the electron source for this PDPWA scheme based on the existing PHIN photo-injector at CERN. The modifications to the existing design, preliminary beam dynamics simulations in order to provide the required electron beam are presented in this paper.
\end{abstract}
\end{frontmatter}
\section{Introduction}
The baseline design specifications of the PHIN photo-injector is given in Table \ref{tbl:phin}. PHIN was designed \cite{phin_design} and commissioned \cite{phin_thesis} to serve for the CLIC drive beam providing a long bunch train with high charge and most of all with a high intensity stability of less than 0.25$\%$ \cite{phin_prstab}. This paper reports on the modifications and parameter adjustments to 'tune' the PHIN photo-injector in order to produce an electron beam compatible with AWAKE project's witness electron beam.
\begin{figure}[!htb]
   \centering
   \includegraphics*[width=87mm]{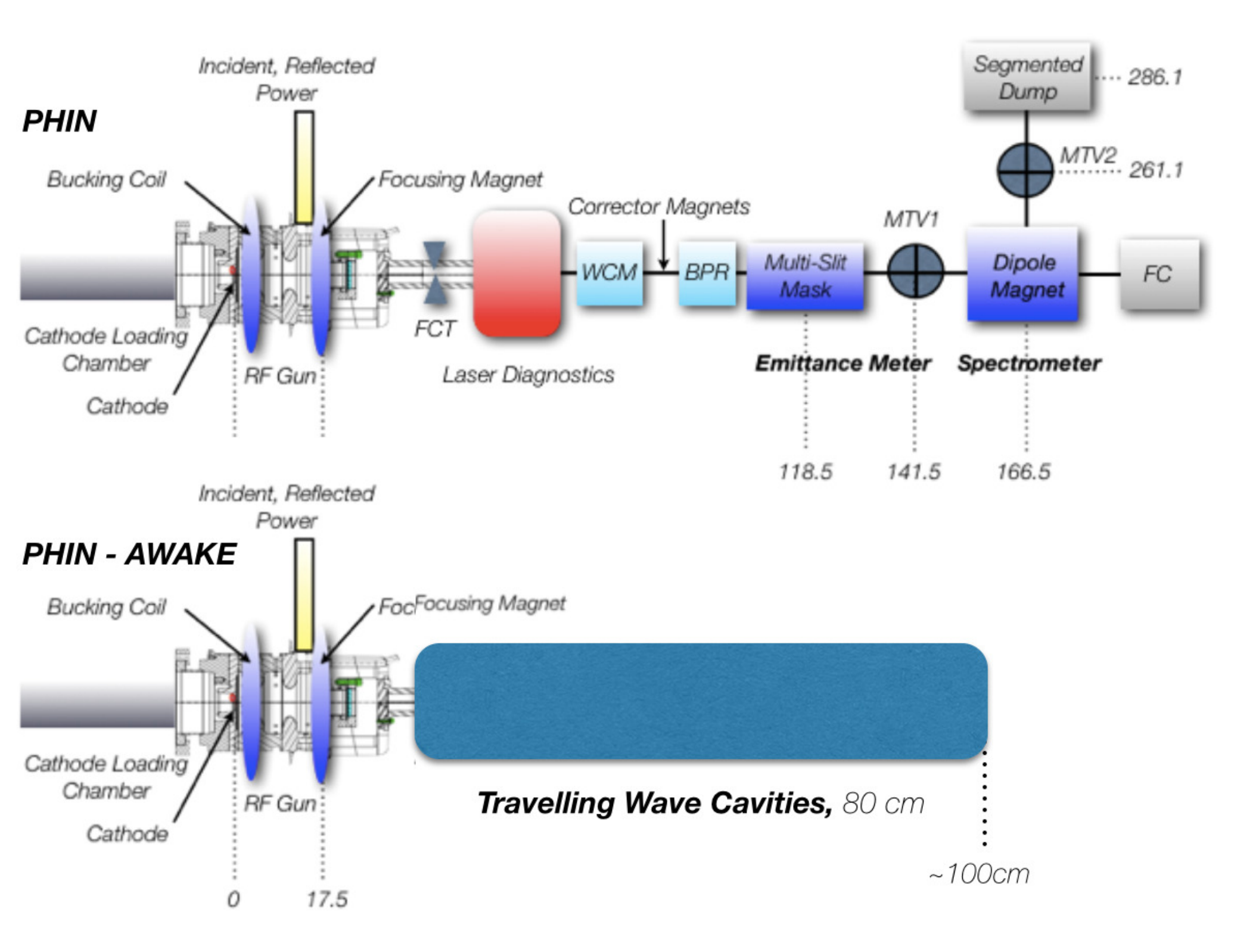}
   \caption{Layout of the baseline and modified PHIN setups.}
   \label{fig:layout}
\end{figure}
\begin{table}[hbt!]
   \centering
   \caption{Baseline Specifications of the PHIN facility.}
   \begin{tabular}{lc}
       \toprule
       \textbf{Parameter} & \textbf{Specification}                     \\
       \midrule
           \bf{Laser}                                    &                  \\
           UV Laser Pulse Energy (nJ)                      & 370  \\
           Micropulse Repetition Rate  (GHz)               & 1.5                 \\
           Macropulse Repetition Rate (Hz)                     &1-5                  \\
           Train Length (ns)     & 1273                    \\
           \bf{Electron Beam}                                    &                  \\
           Charge per Bunch (nC)            &2.33                       \\   
           Charge per Train (nC)            &4446                    \\    
           Current (A)             &3.5                                    \\  
           Norm. Emittance (mm$\,$mrad)            & $<$25                              \\   
           Energy Spread ($\%$)           & $<$1                                     \\   
           Charge Stability ($\%$, rms)           & $<$0.25                                     \\   
           \bf{RF Gun}                                    &                  \\
           RF Gradient  (MV/m)         &85                       \\   
           RF Frequency  (GHz)         &2.99855                       \\   
           Cathode         &Cs$_2$Te                       \\   
           Quantum effficiency ($\%$)        &3                      \\   
       \bottomrule
   \end{tabular}
   \label{tbl:phin}
\end{table}
\section{Beam Dynamics Simulations}
Beam dynamics studies were performed using PARMELA code \cite{parmela}. 
\subsection{Acceleration}
The energy output of the PHIN photo-injector is 5$\,$MeV whereas higher energies are required for the AWAKE witness beam during the injection into the plasma. In order to boost the energy of the electron beam a 80$\,$cm travelling wave structure (TWS) was introduced in the original PHIN design before the exiting diagnostics section (fig.\ref{fig:layout}) on the setup.
\begin{figure}[!htb] 
   \centering
   \includegraphics*[width=85mm]{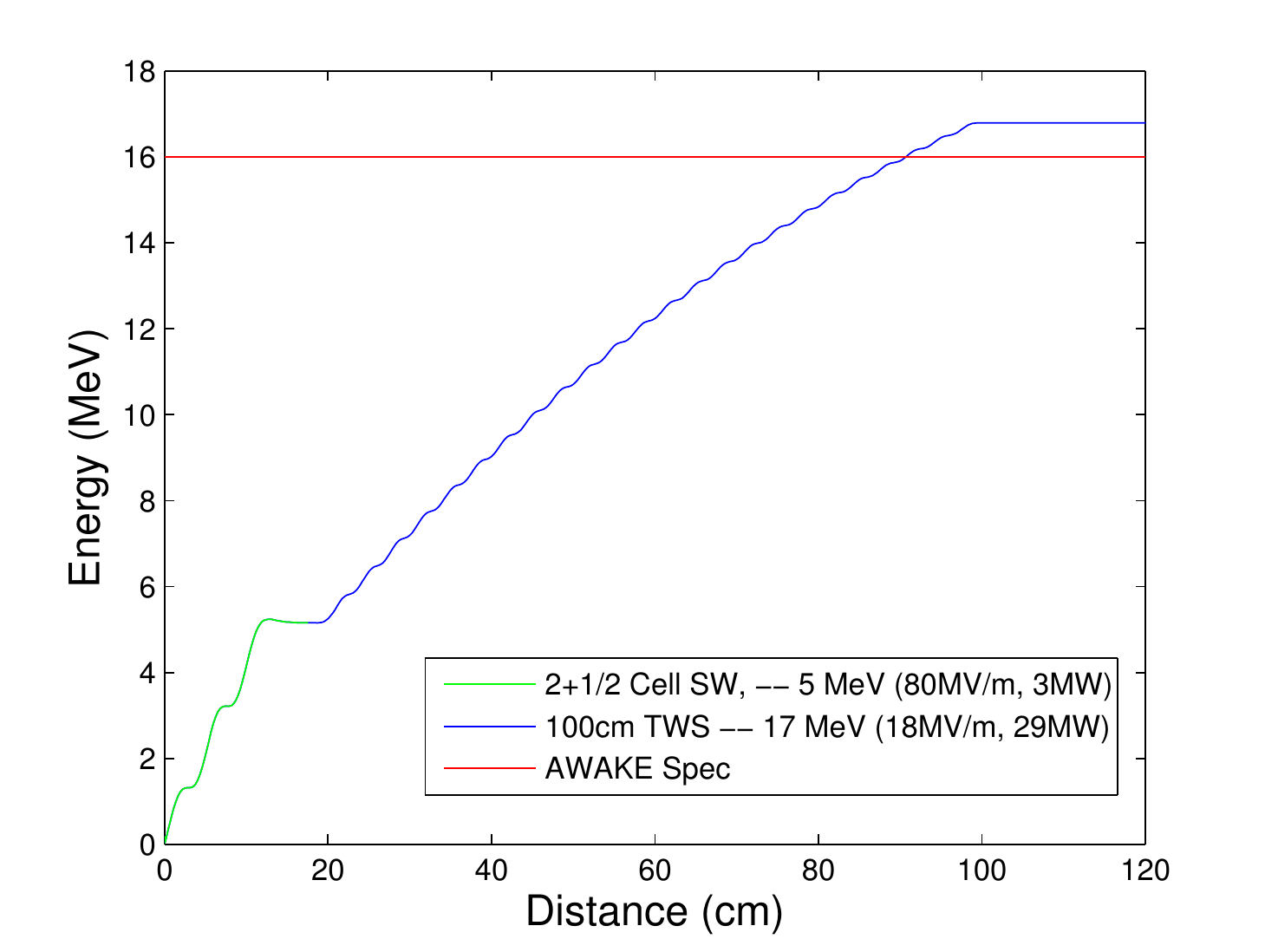}
   \caption{Evolution of beam energy along the beam axis through the RF gun and the travelling wave structure.}
   \label{fig:energy}
\end{figure} 
\begin{figure}[!htb] 
   \centering
   \includegraphics*[width=85mm]{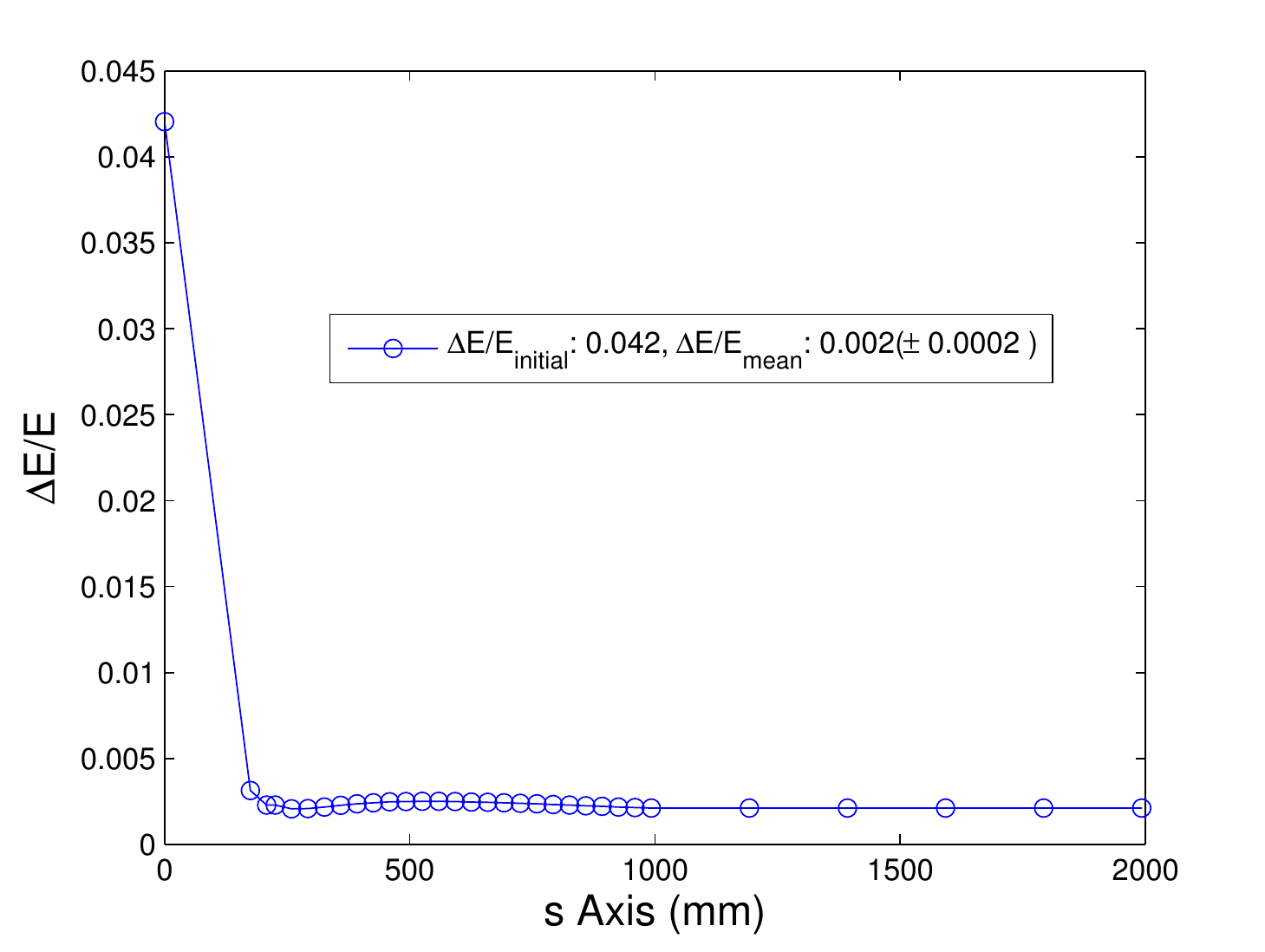}
   \caption{Eenrgy spread along the beam axis.}
   \label{fig:energy_spread}
\end{figure} 
Fig.\ref{fig:energy} presents the energy increase along the beam axis. At the exit of the standing wave cavity beam reaches an energy of about 5$\,$MeV step by step through the cells of the cavity up to 200$\,$mm indicated by the green curve. Energy increase provided by the TWS is shown by the blue curve that reaches a maximum of 17$\,$MeV at about 1000$\,$mm that will be considered as the injection point as far as the optimisation studies are concerned. The evolution of the energy spread during the process is given in fig.\ref{fig:energy_spread} that stabilises at about 200$\,$mm and stays fairly constant at 0.2$\%$.  
\subsection{Emittance} 
Total beam dynamics emittance originates from thermally induced transverse momentum, time dependency of RF field and most dominantly space charge forces in a photo-injector. Due to the use of a laser thermal emittance during particles emerging from the cathode is relatively low and about 4$\%$ of the total emittance whereas the RF induced component is about 13$\%$. The rest of the emittance is due to the space charge forces and the well-known emittance compensation scheme is utilised to minimise the total emittance through the dominant component. In this scheme a pair of solenoid magnets are placed both ends of the RF gun; the solenoid after the gun controls the beam envelope and therefore the emittance where the first one maintains a zero magnetic field at the location of the cathode and prevents particles gaining transverse momentum. 
\begin{figure}[!htb] 
   \centering
   \includegraphics*[width=85mm]{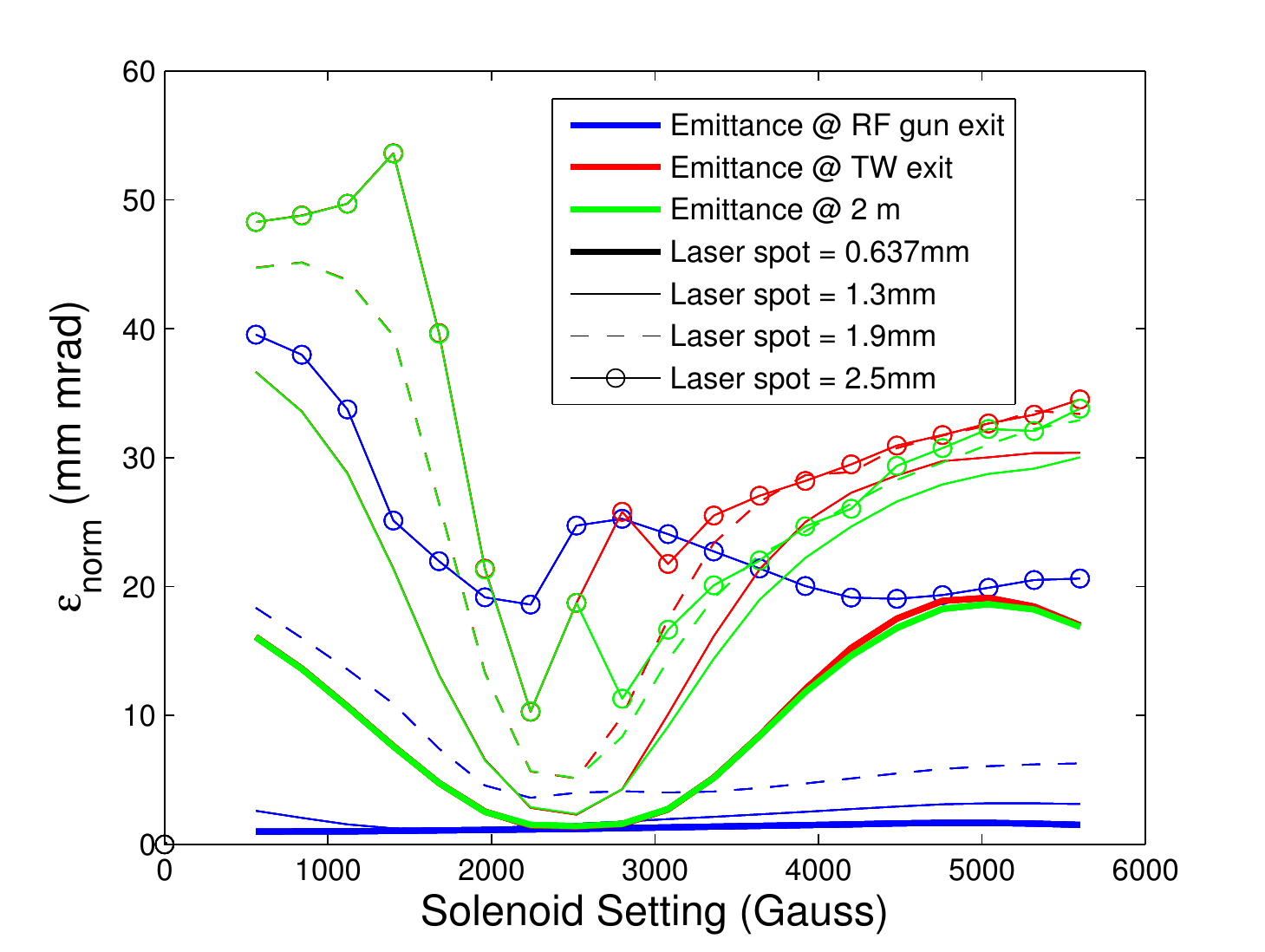}
   \caption{Systematic emittance scan with respect to focusing and laser spot size.}
   \label{fig:laser_solscan}
\end{figure} 
Emittance can be also controlled using the initial conditions such as laser spot size. However one needs to carefully consider while optimising the laser spot size as it effects the charge yield as well. A scan across focusing solenoid field and laser spot size was presented in fig.\ref{fig:laser_solscan} to determine the values of both parameters while producing as much charge as required from the AWAKE electron source. In the figure, colours indicate the longitudinal locations where the emittance is monitored while the plot line-properties correspond to four different laser spot size settings. Consequently, one can easily determine the region for minimum emittance falls within a 2000-3000$\,$Gauss range of the focusing field and emittance increases with the increasing laser spot size. 
\begin{figure}[htb!] 
   \centering
   \includegraphics*[width=85mm]{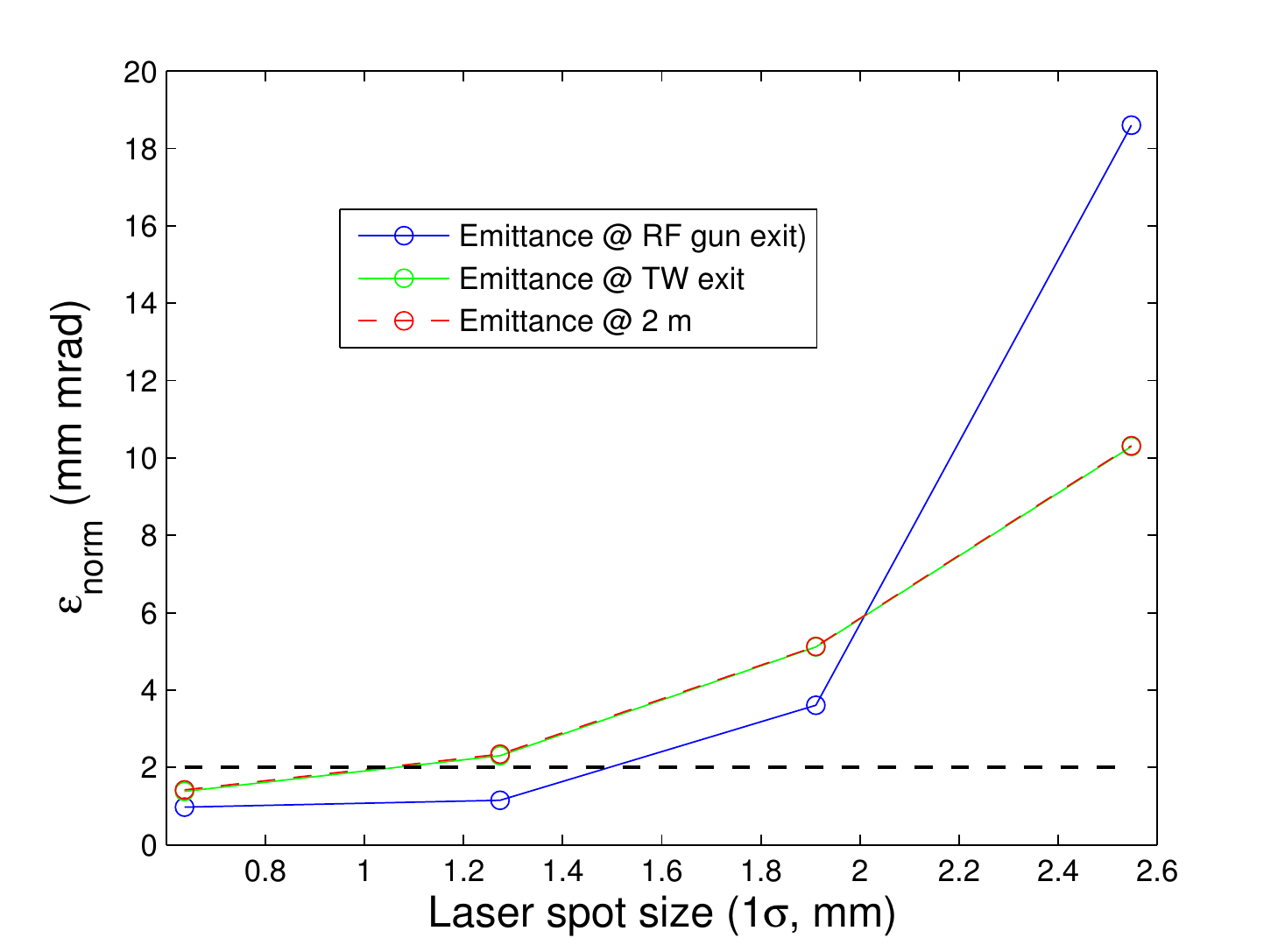}
   \caption{Beam transverse normalised emittance as a function of the laser spot size.}
   \label{fig:laserscan}
\end{figure} 
This last effect can be observed more clearly in fig.\ref{fig:laserscan} where the black line represents the emittance budget determined by the current electron beam transfer line design of the AWAKE project. According to these considerations a focusing field of about 2500$\,$Gauss and a spot size smaller than 1$\,$mm can provide an emittance within the AWAKE electron gun budget.  

The charge production side of the story can be assessed by using the relation in Eq.\ref{eqn:charge} \cite{phin_thesis} where $E_{acc}$ is the accelerating gradient of the standing wave cavity and $\sigma_x$ is the laser spot size. According to this estimation, with a gradient of 80$\,$MV/m laser spot size can have values between 0.2-1$\,$mm while producing 0.2-4.4$\,$C bunch charge and preserving the emittance under 2$\,$mm mrad.
\begin{equation} 
Q_{max}[nC]= \frac{E_{acc}[MV/m]\sigma_x^2[mm^2]}{18}
\label{eqn:charge}
\end{equation}  
\begin{figure}[!htb] 
   \centering
   \includegraphics*[width=84mm]{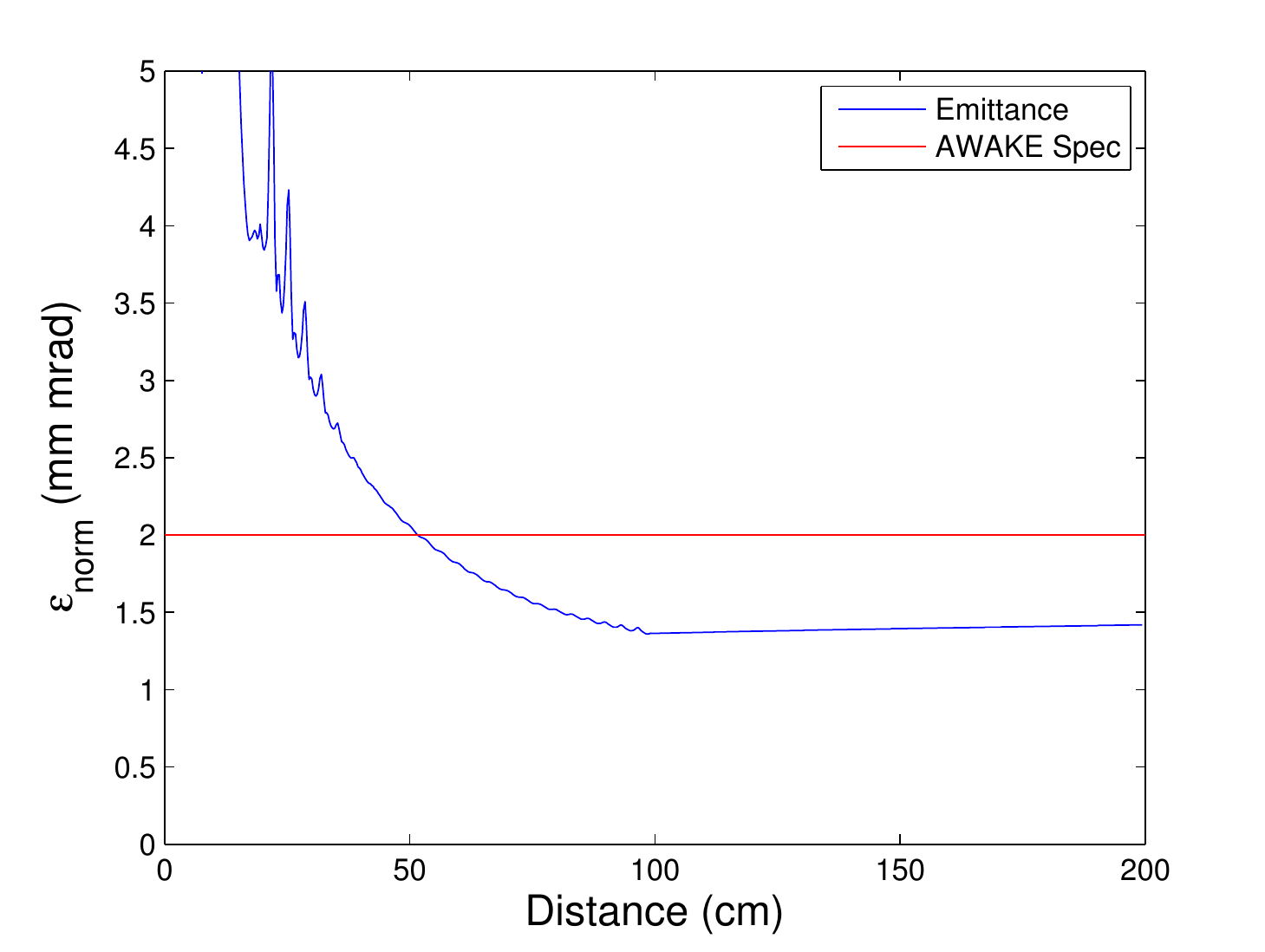}
   \caption{Evolution of the emittance along the beam axis where the solenoids are set for the minimum emittance.}
   \label{emitt_las1}
     \vspace{-1.5em}
\end{figure} 
Emittance evolution along the beamline is presented in fig.\ref{emitt_las1} for the laser spot size of $0.6\,mm$ which provides the minimum emittance after the acceleration section. Emittance oscillates in the standing wave cavity and undergoes adiabatic damping in the travelling wave cavity in the first 100$\,$m.  The deliverable emittance was optimised at the exit of the travelling wave structures and equals to 1.4$\,$mm mrad. 
\begin{figure}[!htb] 
   \centering
   \includegraphics*[width=84mm]{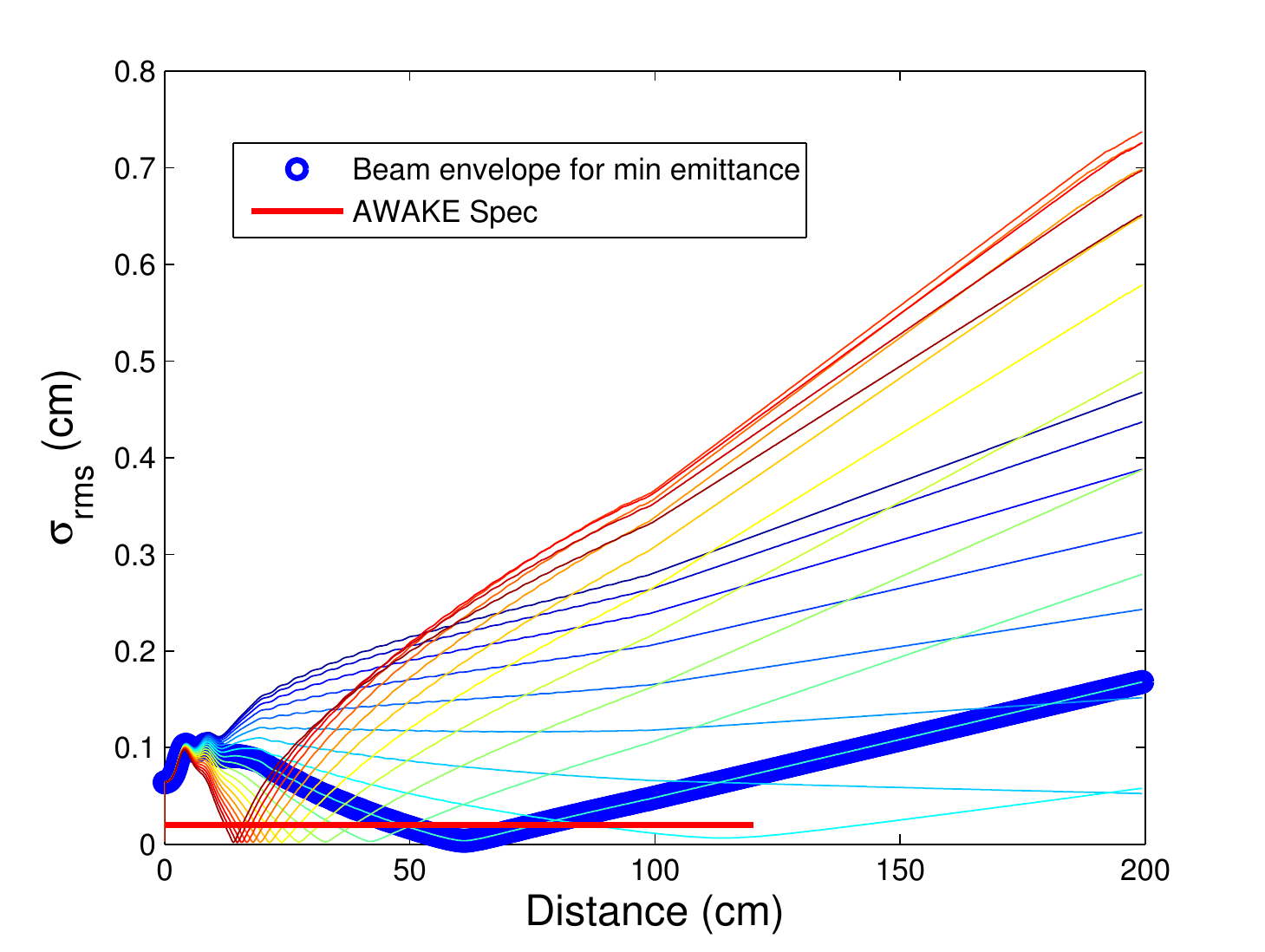}
   \caption{Behaviour of the beam waist along the beam axis as a function of the focusing settings. Dark blue curve indicates the beam envelope during the minimum emittance occurs at the exit of the TWS.}
   \label{fig:sigma_las1}
\end{figure} 
Fig.\ref{fig:sigma_las1} demonstrates how beam waist can be moved along the beam axis by using the two solenoids existing in the setup. The colour-code from blue to red indicates the increasing magnetic field. The dark blue curve emphasises the setting which provides minimum emittance at the exit of the travelling wave structure.
\begin{table}[hbt!] 
   \centering
   \caption{Beam specifications during injection at 100$\,$m.}
   \begin{tabular}{lcc}
       \toprule
       \textbf{Parameter} & \textbf{AWAKE}                      & \textbf{Modified PHIN} \\
       \midrule
           Bunch population                      & 1.25x10$^9$          & 1.25x10$^9$        \\
           Bunch length, mm                     & 2.5           & 0.89       \\
           Bunch radius, mm                     &0.2            & 0.5       \\
           Norm. emittance mm-mrad       & 2            & 1.4        \\
           Energy, MeV                             & 16            & 17       \\
           Energy spread, $\%$                & <1        & 0.2 \\    
       \bottomrule
   \end{tabular}
   \label{tbl:results}
\end{table}
Table \ref{tbl:results} presents the beam specifications required by AWAKE witness beam and the values that can be delivered by modified-PHIN photo-injector. According to the simulation results PHIN photo-injector can be tuned to provide all specifications required by AWAKE project with the exception of the beam radius which can be further optimised by additional focusing.
\section{Conclusions and Outlook}
A systematic study was conducted in order to explore the possibility of adjusting the existing PHIN photo-injector in order to produce the electron beam required by the AWAKE project to be used as the witness beam in the plasma wakefield acceleration scheme. The preliminary results are promising. Therefore a further investigation is worthwhile such as RF phase and amplitude optimisation to maintain acceleration on the RF off-crest. Operation under this condition can provide an energy chirp which might be used to compress the bunch. Furthermore, optimum integration of the accelerating structure should be determined considering engineering and beam dynamics specifications. Focusing elements should be implemented into the setup which will lead re-optimisation of the laser parameters.

\end{document}